\documentclass[9pt,twocolumn,twoside]{revtex4-1}

\usepackage{bm}
\usepackage{csquotes}
\usepackage{mathtools}
\usepackage{textcomp}
\usepackage{amsmath}
\usepackage{amssymb}
\usepackage[utf8]{inputenc}
\usepackage{xcolor}
\usepackage[normalem]{ulem}
\usepackage{amsthm}
\usepackage{enumitem}
\theoremstyle{remark}
\DeclareFontFamily{U}{wncy}{}
\DeclareFontShape{U}{wncy}{m}{n}{<->wncyr10}{}
\DeclareSymbolFont{mcy}{U}{wncy}{m}{n}
\DeclareMathSymbol{\Sha}{\mathord}{mcy}{"58}

\newcommand{\changes}[1]{{\color{black}{#1}}}

\begin{document}

\newcommand{\dname}{demon}
\def\siMD{S1}
\def\siConservation{S2}
\def\siScaling{S3}
\def\siMarkov{S4}
\def\siKinetic{S5}
\def\siNematicKinetic{S6}
\def\siNematicHydro{S7}
\def\siExperiment{S8}
\def\siVideos{S9}

\title{Measurement-Induced Phase Transitions in Informational Active Matter}


%

\author{Bryan VanSaders}
\thanks{Current affiliation: Department of Physics, Drexel University, Philadelphia, PA 19104, USA}
\affiliation{James Franck Institute, The University of Chicago, Chicago, Illinois 60637, USA}

\author{Michel Fruchart}
\affiliation{Gulliver, ESPCI Paris, Université PSL, CNRS, 75005 Paris, France}

\author{Vincenzo Vitelli}
\email{vitelli@uchicago.edu}
\affiliation{James Franck Institute, The University of Chicago, Chicago, Illinois 60637, USA}
\affiliation{Department of Physics, The University of Chicago, Chicago, Illinois 60637, USA}
\affiliation{Kadanoff Center for Theoretical Physics, The University of Chicago, Chicago, Illinois 60637, USA}

\begin{abstract}
\changes{Various} biological and synthetic \changes{media} out of equilibrium can be viewed as \changes{many-ratchet systems} that rectify environmental noise through local measurements and information processing, like in Maxwell's prototypical demon. 
These systems pose a challenge to standard \changes{coarse-graining} approaches because they are better described in terms of decision-making protocols similar to computer programs rather than force laws.
Here, we study a many-body generalization of the Maxwell demon problem: a fluid composed of adaptive particles that achieve collective behavior by biasing noise-driven scattering events \changes{subject to measurements}.
Using a combination of information-theoretic, kinetic, and hydrodynamic tools, we elucidate how microscopic decision-making protocols, rather than microscopic forces, generate macroscopic \changes{active} states \changes{sustained by continuous measurements.
These include} an informational version of flocking \changes{whose order parameter is bounded by the information measured, and the onset of which may be viewed as a measurement-induced phase transition}.
We find that the signature of such microscopic choices is an `informational activity' that selectively compresses phase space, without work, and causes deviations from equilibrium scaling with the magnitude of environmental noise.
We envision applications to noise-induced patterning performed by collections of microrobots \changes{guided by reinforcement learning} or programmable phoretic colloids in \changes{turbulent flows} that\changes{ exploit local measurements and control actions to counteract the scrambling of information by chaos.}
\end{abstract}


\maketitle


A macromolecular machine within a cell and a boat on troubled waters are both subject to relentless random forces \changes{driving chaotic dynamics}: thermal fluctuations in the cell, waves and wind on the sea.
At first sight, these strong fluctuations seem to be a challenge opposing the molecular machine's or boat's goals.
However, both sailors and biological components can exploit noise for their own purposes by sensing their environment, taking actions informed by \changes{local} measurements, and breaking time-reversal symmetry.
This process turns them into \enquote{information engines}~\cite{Feynman2023, Ro2022,horowitz_thermodynamic_2020,   
 obyrne_time_2022, Parrondo2015,Landauer1961}.
Information engines operate not by applying forces but rather by restricting the \changes{volume} of phase space available to the system \changes{through measurements}.
This excluded volume effect leads to entropic forces whose strength is proportional to temperature in a thermal system~\cite{Israelachvili2011,Doi2013,Hanggi1990,Reguera2001,Braun2016,WissnerGross2013,Devereux2023}, like the pressure in an ideal gas~\cite{Doi2013}, elasticity in rubber~\cite{Flory1943}, or the phoretic forces underlying self-propelled colloids~\cite{Moran2017,Brady2010,Sabass2012,Bebon2024}.
Crucially, information engines can {\it actively} change the excluded volume in phase space depending on their state, setting up a feedback mechanism that drives the system out of equilibrium by rectifying noise and \changes{chaotic dynamics} into a desired motion or behavior \changes{through local measurements and information processing}~\cite{Parrondo2015,Buisson2024}.
\changes{This mechanism} is believed to play a key role in biological and behavioral processes~\cite{Hopfer2002,Sartori2015, Andrieux2008,Ito2015,Tu2008, Sartori2014,Binder2011,Boel2019,Mizraji2021,Leff2002,Miller2001,Stephens2019,VanLoon2011,Sachs2005,Braun2016,Charlesworth2019,Heins2024,Devereux2023}.

Considerable research has focused on single-body information engines, in which a lone agent (colloquially known as a Maxwell demon) exploits information as a resource to perform work towards a goal. 
However, theoretical descriptions of many-body information engines are comparatively rare \cite{ziepke2022multi, jung_kinetic_2025, dillavou_demonstration_2022}.
Many natural systems are composed of large numbers of cooperating agents that individually collect, process, and act on information (e.g.~cells in a tissue, ant colonies, bird flocks, and human communities).
Furthermore, over the last two decades, per-particle external feedback control has been demonstrated in synthetic active matter systems, such as programmable phoretic colloids \cite{Ghosh2009, Lavergne2019, Tierno2008, Yang2018, Demirors2018, Fernandez2020, wang2024harnessing, heuthe2024counterfactual, qian2013harnessing, muinos2021reinforcement, cichos2020machine, snezhko2011magnetic}.
As colloidal scale objects are engineered to be more complex \cite{Alvarez2021} and robotic platforms shrunk towards the colloidal regime \cite{Miskin2020, Reynolds2022}, a new category of small-but-capable machines are emerging \cite{Palagi2018, Huang2020, liu_colloidal_2023}.
Yet, \changes{little is known about collective phenomena that arise in such many-body systems as a result of their ability to perform local measurements and control actions.
We henceforth refer to these many-body systems as informational active matter.
In this class of systems, measurements and control actions can counteract the scrambling of information by the underlying chaotic dynamics (e.g.~microscopic collisions). 
As a consequence, this information, typically encoded in low-entropy ordered states, can survive in a nonequilibrium steady-state. 
}

To distill the non-equilibrium \changes{collective} effects that arise purely from informational activity, we introduce a minimal extension of what is perhaps the simplest interacting many-body system, the hard disk gas.
This extension, amenable to theoretical treatment, explicitly accounts for the feedback control processes whereby each agent (i.e.~Maxwell demon) makes control decisions \changes{informed by} the outcome of measurements.
Such a gas of Maxwell demons can reach through consensus (but without exerting any work) a nematic flocking-like state that we \changes{analyze}, from kinetic to hydrodynamic theory, by coarse-graining microscopic decision-making policies \changes{informed by measurements,} rather than pre-determined microscopic forces.
\changes{We show that the steady-state magnitude of the nematic-flocking order parameter is bounded by the information measured, which suggests viewing the onset of informational flocking as a phase transition induced by measurements \cite{Jin2022,Gerbino2025,Pizzi2022,Willsher2022,Pizzi2024}.
This classical mechanism is reminiscent of quantum measurement-induced phase transitions (MIPT), where quantum measurements counteract the scrambling of information originating from the underlying unitary dynamics \cite{Skinner2019,Choi2020,annurev}.}

\noindent \textbf{The \dname\ gas model.}
Consider a gas of hard disks at finite temperature, undergoing random collisions with each other.
The velocities of individual particles follow the Maxwell-Boltzmann (MB) distribution.
Imagine now that these disks could \enquote{opt out} of collisions with other disks: this would allow them to select whatever velocity is desired by simply waiting until they obtain it randomly by collision, then subsequently avoiding scattering interactions.
Such a system would act as a distributed version of the demon imagined by Maxwell, that opens and closes a trapdoor to create a temperature difference between two sides of a vessel filled with gas~\cite{Maxwell1871,Thomson1874,Landauer1961,Szilard1929, Plenio2001, Leff2002}.

\begin{figure*}[ht]
\centering
\includegraphics[width=\textwidth,keepaspectratio]{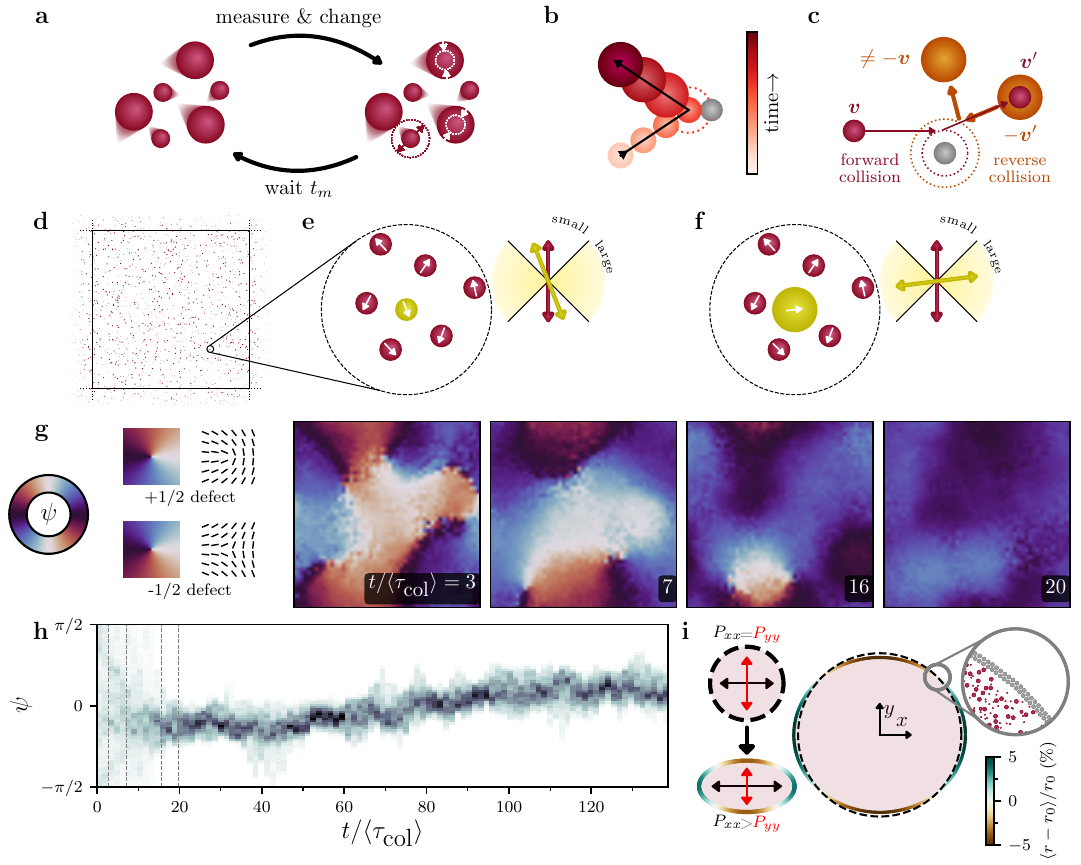}
\caption{\textbf{The \dname\ gas uses measurement to break detailed balance without work.}
\textbf{a}. A gas of feedback controllers, hard disks capable of changing their diameter in response to a measurement.
Here we consider gases where measurements are instantaneous and synchronized, with a delay of $t_m$ between subsequent measurements.
\textbf{b}. Controllers maintain energy conservation during collision events by disallowing diameter changes that would introduce particle overlaps.
\textbf{c}. Diameter as a function of particle velocity breaks detailed balance - a `reverse' collision that begins where a `forward' collision ends does not recover the initial configuration of the forward collision.
\textbf{d}. A \dname\ gas with periodic boundary conditions, simulated via molecular dynamics.
\textbf{e}-\textbf{f}. Diameter changing conditions for a nematically-aligning \dname\ gas.
\textbf{e}. When particle velocity is aligned (or anti-aligned) with the majority of nearby neighbors, a small diameter is chosen.
\textbf{f}. When particle velocity is perpendicular to the majority of nearby neighbors, a large diameter is chosen.
\textbf{g}. Snapshots of binned orientation fields for an initially passive gas at short times after diameter rescaling is enabled.
$\langle \tau_{col}\rangle$ is the mean collision time assuming an even split of large and small diameter particles.
Nematic defects form and annihilate leading to a nearly uniform orientation.
\textbf{h}. Long-time distribution of particle orientations for the gas in part (g). 
The times corresponding to snapshots in (g) are indicated with dashed lines.
\textbf{i} The anisotropic pressure of a \dname\ gas trapped within a passive flexible container inflates it into an ellipsoidal shape (colored by radial deformation $r$ from circular initial condition $r_0$).
}\label{fig:nematic_intro}
\end{figure*}

Here, we implement such a \dname\ gas by endowing every disk in the gas with the ability to change its size, and therefore its scattering cross-section, depending on its environment.
As shown in Fig.~\ref{fig:nematic_intro}a, every period $t_m$, every particle $i$ measures its current state (position $\bm{x}^{(i)}$ and velocity $\bm{v}^{(i)}$) as well as its environment $\mathcal{E}^{(i)}$ (i.e.~some set of information about the current state of nearby particles).
Based on this data, it changes its size according to a choice function $D(\bm{x}^{(i)}, \bm{v}^{(i)}, \mathcal{E}^{(i)})$ giving the new diameter of the particle.  

In order to bring the informational aspects to the fore, we focus on the less explored limit where the functions $D$ do not result in particles exerting work on each other.
In most experimental contexts, the informational (i.e.~entropic) effects highlighted here will coexist with active forces of enthalpic origin on which active matter theories typically focus \cite{Alvarez2021,Ghosh2009, Lavergne2019, Tierno2008, Yang2018, Demirors2018, Fernandez2020}.
In the limit of vanishing work, size changes are confined to the times between collisions, i.e.~when such a change would not cause overlaps (Fig.~\ref{fig:nematic_intro}b), requiring that $\mathcal{E}^{(i)}$ provides proximity information about neighbors.
As a consequence, the quantities conserved in a standard hard disk gas (mass, linear momentum, total kinetic energy) are also conserved in the \dname\ gas (see SI section \siConservation).

Nonetheless, coupling particle size and velocity drives the system out of equilibrium, as shown in Fig.~\ref{fig:nematic_intro}c.
Crucially, collisions are no longer \enquote{undone} by reversing particle velocities if sizes also change, and hence detailed balance is broken (a hallmark of active systems, but here achieved with energy and momentum conservation intact).
As we now show, specifying the choice function $D$ allows us to control the behavior of the fluid.

\noindent \textbf{Informational flocking.} We start by implementing an informational version of flocking.
Flocking is typically modeled as arising from interaction torques that bring the orientations of self-propelled particles into alignment.
No such forces are present in our model, so flocking is instead produced by {\it local} control actions, i.e., the particles (agents) need not know their global positions.
The particles simply align their velocities with their neighbors by selectively reducing their size when they are aligned and increasing their size when they are not aligned.
\changes{Over many collision events, this selective biasing of scattering cross-section results in aligned groups.}
As the total linear momentum is conserved, it is not possible to change the average velocity in the system (see below for a variant where this constraint is lifted), so we have a {nematic} version of flocking in which particles tend to align their velocities $\bm{v}^{(i)}$ modulo reflection $\bm{v}^{(i)} \to - \bm{v}^{(i)}$.
This is represented in Fig.~\ref{fig:nematic_intro}d-f: the current particle (in yellow) computes the average direction of neighboring particles (red) and compares it to its own. 
Depending on whether the double-sided arrows are aligned or not (white and shaded regions in panels e-f), the particle adopts (between collisions) a small diameter $D_{\text{S}}$ or a large diameter $D_{\text{L}}$.
Each particle takes a diameter $D(\theta) = D_{\text{S}} + \Delta D \; \mathcal{H}( \cos^2\theta - 1/2)$
where $\mathcal{H}$ is the Heaviside step function, $\Delta D=D_{\text{L}}-D_{\text{S}} \geq 0$ is the diameter difference,
and $\theta$ is the angle between the particle's direction of motion and the local average direction of motion in a region of volume $\Omega$ around the current particle (Fig.~\ref{fig:nematic_intro}e-f).
Formally, the average direction of motion is determined as the major axis of the local pressure tensor
$P_{ij} = - \rho \langle p_i p_j \rangle_\Omega$, in which $\bm{p}\equiv\bm{v}-\bm{u}_\Omega$, $\bm{u}_\Omega\equiv\langle \bm{v} \rangle_\Omega$, and $\rho$ is the gas density.
Indeed, the traceless part $Q_{ij} = P_{ij} - (P_{kk}/d) \delta_{ij}$ of the (symmetric) pressure tensor can be interpreted as a nematic order parameter and decomposed as $Q_{ij} = 2 Q [ n_i(\psi) n_j(\psi) - \delta_{ij}/d]$ where $d$ is the space dimension, $\hat{n}(\psi)$ is a unit vector making an angle $\psi$ with a fixed direction (e.g.~the $x$ axis), and $Q^2\equiv ||\bm{Q}||^2 \equiv Q_{ij}Q_{ji}/2\ge 0$.
Note that here, the symmetric traceless tensor $\bm{Q}$ describes the motion of the particles, in contrast with nematic liquid crystals where the nematic tensor describes their orientation.

Figure~\ref{fig:nematic_intro}g-h show the result of molecular dynamics (MD) simulations of this model with periodic boundary conditions (see methods and SI section \siMD\ for details).
We observe that the gas acquires a finite nematic order parameter $\bm{Q}$ starting from a disordered initial configuration (the equilibrium state of the passive gas) as the rotation symmetry of the system is spontaneously broken, see panel \ref{fig:nematic_intro}g for a plot of the local orientation $\psi=\arctan(Q_{12},Q_{11})/2$ (modulo $\pi$) and panel \ref{fig:nematic_intro}h for the time evolution of the histogram of $\psi$. 
Nematic defects in the orientation field eventually relax by annihilation (Fig.~\ref{fig:nematic_intro}g), leaving the system with a single dominant (but thermally fluctuating) orientation.
Note that these simulations are too small to prove the emergence of true global orientational order \cite{chate2024dynamic, mahault2019quantitative}, and we leave the development of a fluctuation theory and accompanying scaling simulations to future work (see SI section \siScaling).

This alignment is accompanied by a stable anisotropy in the system-wide pressure tensor.
Fixed diameter disks, in contrast, can only exhibit an isotropic pressure tensor.
This ability can be harnessed to perform work. 
For instance, the gas can asymmetrically inflate a symmetric balloon, as shown in Fig.~\ref{fig:nematic_intro}i. 
The ellipsoidal shape of the balloon arises from the interplay between its surface tension (that favors a symmetric shape) and the nonequilibrium anisotropic pressure tensor of the gas (that favors an asymmetric shape).
Such a mechanism could for instance be exploited to actuate a synthetic membrane from the inside.

\noindent \textbf{Thermodynamics of information-driven fluids.}
In order to understand how the model in Fig.~\ref{fig:nematic_intro} harnesses information to maintain a non-equilibrium steady-state and to perform work on a load, we turn to the framework of information thermodynamics~\cite{Parrondo2015,Cao2009,horowitz2010nonequilibrium, esposito_second_2011}.
In a nutshell, this framework provides a formulation of the second law of thermodynamics that integrates the fact that the entropy of a system can be reduced by acquiring information about it, and subsequently driven out of equilibrium by using control actions conditioned on the measurement results.
To do so, one first introduces the non-equilibrium free energy $\mathcal{F}[\rho] = \langle U \rangle_\rho -  T \, S[\rho]$ associated to a (possibly non-equilibrium) statistical state represented by a probability distribution $\rho$,
where $\langle U\rangle_\rho$ is the mean internal energy and $S[\rho]$ the entropy of the distribution. 
\changes{In general, the joint distribution of all measured variables must be considered to compute entropy.
Here, we restrict ourselves to the single-particle velocity distribution function (a mean field approximation assuming that the alignment of a local neighborhood changes much slower than the individual velocities within it).}
In terms of these quantities, our assumption that the particles are not permitted to exert work on each other means that the mean internal energy $\langle U\rangle_\rho$ is constant.
The second law of thermodynamics for feedback processes then states that the change $\Delta \mathcal{F}$ between two states satisfies~\cite{esposito_second_2011, Parrondo2015}
\begin{equation}
    \label{secondlaw}
    \Delta \mathcal{F} \leq W + k_{\text{B}} \, T \, I
\end{equation}
where $W$ is the work done on the system during the process, and $I$ the mutual information between the microstates of the system and the results of the measurements used for the feedback control.

\begin{figure*}[htbp]
\centering
\includegraphics[width=\textwidth,keepaspectratio]{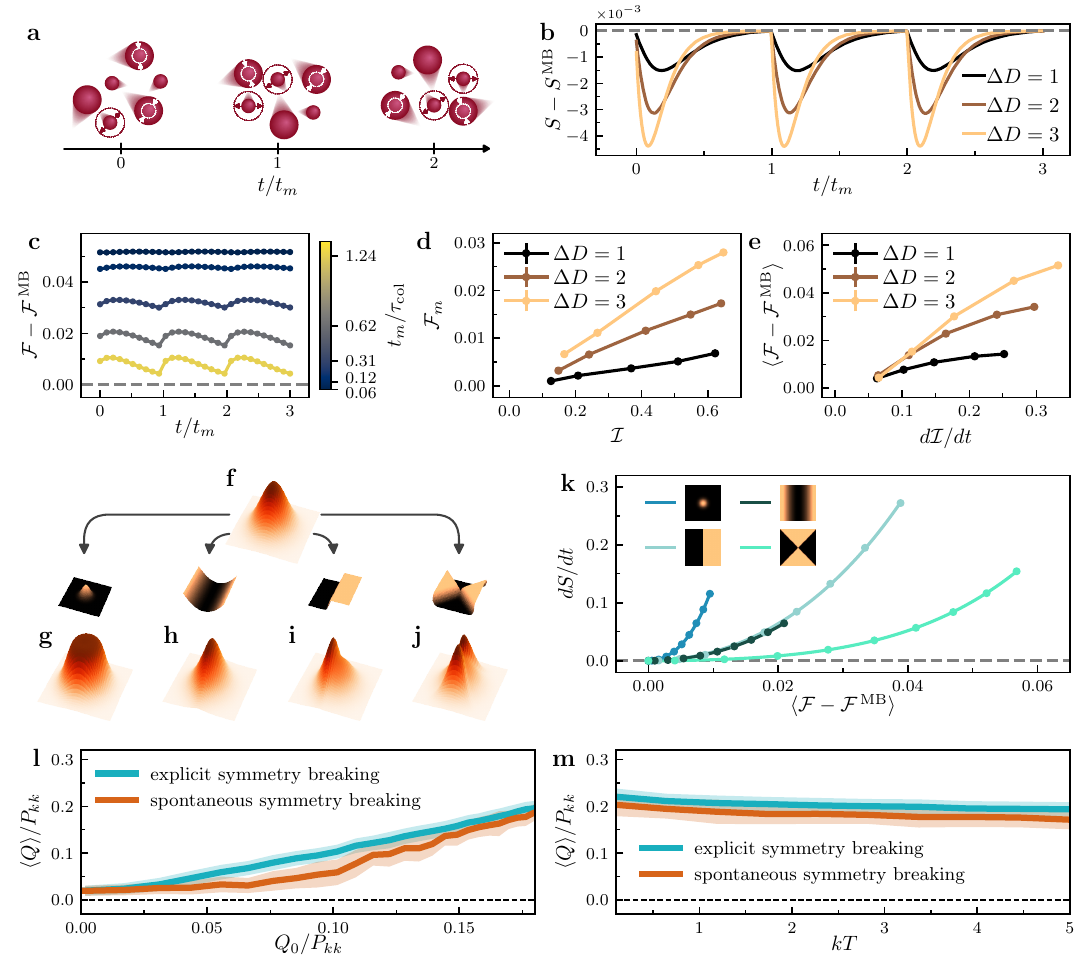}
\caption{\textbf{The \dname\ gas is driven by information}.
\textbf{a}. Particles instantaneously measure their velocity and update their diameters (subject to constraints due to proximity) at an interval of $t_m$.
\textbf{b}. The entropy (relative to the Maxwell-Boltzmann distribution) of a \dname\ gas measured from its velocity distribution function.
Here the measurement interval is several times longer than the mean collision time ($t_m/\tau_{\text{col}}=[3.6,4.3,5]$ for $\Delta D=[1,2,3]$).
\textbf{c}. Relative entropy of \dname\ gases with a range of measurement intervals, up to approximately one collision time.
\textbf{d}. The nonequilibrium free energy injected at each measurement, as estimated by fitting curves similar to (c) as a function of the information per measurement of a Markov process describing the diameter of particles (see extended Fig.~\ref{efig:parameter_fitting}).
\changes{Note that $I$ increases with the interval between measurements $t_m$.}
\textbf{e}. The average nonequilibrium free energy as a function of the information rate of the diameter Markov process.
\changes{Note that for the values studied here $dI/dt$ increases as $t_m$ is reduced.}
\textbf{f}-\textbf{j}. The result of various diameter functions on the gas velocity distribution.
\textbf{f}. The Maxwell-Boltzmann distribution in the velocity plane (constant diameter).
\textbf{g}. A zero-centered Gaussian diameter function.
\textbf{h}. A quadratic diameter function of one velocity component.
\textbf{i}. A diameter step function that breaks reflection symmetry.
\textbf{j}. A diameter step function in angular velocity coordinates that breaks pressure symmetry.
\textbf{k}. The entropy production rate of various \dname\ gases as a function of non-equilibrium free energy.
\textbf{l}. Growth of the nematic order parameter with $Q_0$ for simulations of \dname\ gases with explicitly (blue) and spontaneously (orange) broken symmetry.
The system-wide average of the order parameter does not approach zero as $Q_0\to 0$ due to thermal fluctuations.
\textbf{m}. Nematic order parameter as temperature is increased 50-fold.
Nematic ordering is approximately independent of temperature.
}\label{fig:info_hydro}
\end{figure*}

The quantity $k_{\text{B}} I$ corresponds to the entropy reduction due to the control actions.
The mutual information $I$ can easily be computed in simple situations, but in general its evaluation is a challenging task, in particular because of correlations between successive measurements~\cite{Parrondo2015,Cao2009,esposito_second_2011,Horowitz2010,Sagawa2010}.
In our case, all measurements are synchronized (Fig.~\ref{fig:info_hydro}a), but they take time to influence observable gas properties, because their consequences are only felt during collisions.
Hence, their effect is smeared out over the average time between collisions $\tau_{\text{col}}$.
Once a particle has collided, its previous measurement becomes outdated because it may have changed direction, so the next collision will tend to equilibrate the gas.
After a few $\tau_{\text{col}}$ we expect the system to behave like an equilibrium hard disk gas until the next measurement.
Figure \ref{fig:info_hydro}b shows the reduction in entropy relative to the MB distribution $S - S^{\text{MB}}$ over time for $t_m \gg \tau_{\text{col}}$,
showing successive dips in the entropy followed by a relaxation to equilibrium.
Note that energy conservation implies that $\mathcal{F}-\mathcal{F}^\text{MB} = -T(S-S^\text{MB})$.
In fig.~\ref{fig:info_hydro}c, $t_m$ is reduced, and the peaks corresponding to entropy dips are less and less pronounced until a stationary balance between entropy reduction and relaxation towards thermodynamic equilibrium is observed (blue curves). 
When $t_m \lesssim \tau_{\text{col}}$, the system does not fully relax to the equilibrium MB state, inducing correlations between subsequent measurements.

In order to assess the effect of measurements, we therefore take a phenomenological approach that captures the unidirectional coupling between the information-carrying degree of freedom (the particle diameters) and the velocities of the particles.
We model the effect of each measurement as an instantaneous change of the free energy $\mathcal{F}_{\text{ic}}$ of the information carriers by a value $\mathcal{F}_{\text{m}}$, which takes place every $t_m$.
The free energy stored in the diameters is purely informational (entropic). 
It relaxes to its equilibrium (Maxwell-Boltzmann) value $\mathcal{F}_{\text{ic}}^{\text{MB}}$ through collisions over a relaxation timescale $\tau$, each time it is used to make a decision, and this leads to an equal increase in the free energy $\mathcal{F}$ associated with the velocities of the particles. 
Finally, $\mathcal{F}$ also relaxes to its equilibrium value $\mathcal{F}^{\text{MB}}$ with the same collisional time scale $\tau$, in the same way as in a normal hard disk gas.
This translates into the equations of motion
\begin{subequations}
\begin{align}
    \dot{\mathcal{F}}_{\text{ic}} &= - \frac{1}{\tau} [\mathcal{F}_{\text{ic}} - \mathcal{F}_{\text{ic}}^{\text{MB}}] + \mathcal{F}_{\text{m}} \Sha_{t_m}(t) \\
\intertext{and}
    \dot{\mathcal{F}} &= -\dot{\mathcal{F}_{\text{ic}}} - \frac{1}{\tau} [\mathcal{F} - \mathcal{F}^{\text{MB}}] \label{neq_fe_eom}
\end{align}
\end{subequations}
in which $\Sha_{t_m}$ represents a series of unit impulse trains spaced by $t_m$.
In the steady-state, the departure $\Delta\mathcal{F}^*$ of $\mathcal{F}$ from the Maxwell-Boltzmann value just before a measurement does not change from a period to the next and one can show that $\Delta\mathcal{F}^* \propto \mathcal{F}_{\text{m}}$~\footnote{
The steady-state evolution over one period $0 \leq t \leq t_m$ is then $\mathcal{F}_{\text{ic}}(t) = \mathcal{F}_{\text{ic}}^{\text{MB}} + (\mathcal{F}_m + \Delta\mathcal{F}^*) e^{-{t}/{\tau}}$ and 
$\mathcal{F}(t) = \mathcal{F}^{\text{MB}} + (\Delta\mathcal{F}^* + \left(\mathcal{F}_m + \Delta\mathcal{F}^*\right) t/\tau )e^{-{t}/{\tau}}$.
Imposing that $\mathcal{F}(0^-) = \mathcal{F}^{\text{MB}} + \Delta\mathcal{F}^*$ equals $\mathcal{F}(t_m^-) = \mathcal{F}^{\text{MB}} + (\mathcal{F}_m + \Delta\mathcal{F}^*) e^{-{t_m}/{\tau}}$ yields $\Delta\mathcal{F}^* = \mathcal{F}_m/T^*$, where $T^*=(e^{t_m/\tau}-1)\tau/t_m - 1$.
Combining expressions, the average value over a period is $\langle \mathcal{F}\rangle_{0\to t_m} = \mathcal{F}_m(2/T^*+1)(-e^{-t_m/\tau}(t_m+\tau)+\tau)/t_m+\mathcal{F}^\text{MB}$.
}.
Fitting to numerical free energy data shown in Fig.~\ref{fig:info_hydro}c provides estimates of the parameters $\mathcal{F}_m$ and $\tau$ (see methods and extended Fig.~\ref{efig:parameter_fitting}).

To estimate the amount of information collected by the disks, we describe the binary choice by a two-state Markov process, where the two states correspond to the two possible sizes a particle can take (see SI section \siMD\ for detailed methods).
This neglects correlations between particles, but captures correlations between successive measurements.
The transition probabilities $T_{ij}$ and steady-state distribution $\mu_i$ of the Markov chain are estimated from the numerical simulations, yielding the mean information per step as \cite{Cover2006, Cao2009, Huang2020},
\begin{equation}
I \gtrsim I_{\text{MC}} \equiv -\sum_{ij} \mu_i \, T_{ij} \, \ln T_{ij}. \label{eq:markov_entropy}
\end{equation}
Strictly speaking, this is only a lower bound on the total information per step $I$.
The results are shown in Fig.~\ref{fig:info_hydro}d, where we observe that the free energy impulses $\mathcal{F}_m$ scale with and are bounded by the Markov estimate of information gained per measurement.

Due to correlations between measurements the amount of information gained per measurement decreases when $t_m<\tau_{\text{col}}$.
However the average non-equilibrium free energy $\langle\mathcal{F}-\mathcal{F}^{\text{MB}}\rangle$ increases with the rate of information acquisition (i.e.~for rapid correlated measurements).
\changes{For values of $t_m$ studied here, increasing the rate of measurement increased $dI/dt$ (i.e.~correlations do not entirely negate the advantage of more rapid acquisition).}
Fig.~\ref{fig:info_hydro}e shows that average non-equilibrium free energy scales with the rate of information gain.
Overall, a larger diameter contrast $\Delta D$ leads to a larger effect (panels b and d-e), while measurement time determines the mean and excursion of $\mathcal{F}-\mathcal{F}^\text{MB}$.
Similar arguments can be used to obtain estimates of free energy expenditure in the case of a single \dname\ particle immersed in a passive thermalized gas, see extended Fig.~\ref{efig:markov} and SI section \siMarkov.

By analogy with eq.~\eqref{secondlaw}, we expect $\Delta\mathcal{F}^* \propto \mathcal{F}_{\text{m}} \leq k_{\text{B}} T  I$. 
This bound on the deviation from equilibrium shows to what extent the measurement and feedback control process drives informational active matter out of equilibrium.
In order to assess what nonequilibrium states are reached and how they evolve, we focus on the steady-state limit ($t_m/\tau_\textrm{col} \ll 1$) and turn to a kinetic theory that includes the effects of active phase space compression.

\noindent \textbf{Kinetic theory with compressible phase space.}
Within the framework of kinetic theory~\cite{Kardar2007,Dorfman2021,Cercignani1994} we start with the Boltzmann equation,
\begin{equation}
\partial_t f  + \bm{v} \cdot \nabla_{\bm{x}} f + \bm{F} \cdot \nabla_{\bm{v}} f = \mathcal{C}[f]\label{eq:boltzmann_equation}
\end{equation}
\noindent where $f(t,\bm{x},\bm{v})$ is a single particle velocity distribution function, $\bm{F}$ represents external forces (that we neglect here), and $\mathcal{C}$ is the collision operator, which describes the manner in which particle collisions redistribute probability density in phase space.
To account for the diameter feedback control process, we consider a modified version of the usual Boltzmann quadratic collision operator
\begin{equation}
\mathcal{C}[f] = \int (g_+ - g_-) d \sigma d\bm{v}_* 
\label{eq:col_op}
\end{equation}
\noindent 
in which
$g_+ = \alpha(\bm v, \bm v', \bm v_*, \bm v_*') B(\sigma,\Delta \bm{v})  f(\bm{v}')f(\bm{v}'_*)/2$
and $g_- = B(\sigma,\Delta \bm{v}) f(\bm{v})f(\bm{v}_*)/2$, 
where $\bm{v}$ and $\bm{v}_*$ are the velocities of particles undergoing collision, $\bm{v}'$ and $\bm{v}'_*$ are their post-collision velocities, and $\Delta\bm{v}=\bm{v}-\bm{v}_*$.
The collision kernel $B=B(\sigma,\Delta \bm{v})$ describes the scattering of particles for a given impact parameter $\sigma$.

Crucially, the feedback-controlled bias of transitions between velocity states introduces distortions of phase space represented by an $\alpha(\bm v, \bm v', \bm v_*, \bm v_*') \ne 1$ term.
When detailed balance holds, $\alpha = 1$ and the steady-state solution of Eq.~\ref{eq:boltzmann_equation} is the Maxwell-Boltzmann distribution.
This is not the case with the \dname\ gas, where $\alpha$ is generically a function of measured microscopic properties of the system, e.g.~velocities.
Dissipation in granular media \cite{Brilliantov2004} or mechanical self-propulsion in flocking media \cite{Bertin2006} can also introduce such compressible phase space flows.
In particular, the alignment rules in flocking can be interpreted as the result of information processing and decision making at the level of the agents~\cite{Ferretti2022,Geiss2022,Attanasi2014,Cavagna2013,Ren2018,Devereux2023}. 
In usual models of flocking like the Vicsek model, however, it is difficult to keep track of information exchanges and to disentangle the fully informational (entropic) part from the part requiring mechanical work.

For the \dname\ gas, the collision operator should describe the evolution of a variable-diameter hard disk gas.
Therefore the collision kernel $B$ takes on the form for hard disk collisions (see methods and extended Fig.~\ref{efig:microreversibility}), while $\alpha$ is a ratio of collision cross sections before and after the collision.
For homogeneous \dname\ gases, \changes{with diameter functions of velocity only,} $\alpha$ can be shown to have the form (see methods)
\begin{equation}
\alpha = \frac{D(\bm{v}')+D(\bm{v}_*')}{D(\bm{v})+D(\bm{v}_*)}. \label{eq:alpha}
\end{equation}
\changes{Only treating diameter as a function of single-particle velocities greatly simplifies the problem but requires other inputs to the diameter function (such as the average local velocity considered in the flocking example above) to be treated as constants.
This amounts to requiring a separation of timescales, with single-particle velocities assumed to evolve much faster than the average of their local neighborhood.}
In the SI (sections \siKinetic\ and \siNematicKinetic), we show that the steady-state distribution function can be written as $f^d=(1+\mathcal{D})f^\textrm{MB}$ as a function of the Maxwell-Boltzmann distribution $f^\textrm{MB}$ and a function $\mathcal{D}(\bm{v})$ that preserves the conservation of particle number, linear momentum, and kinetic energy.
\changes{The non-equilibrium distribution $f^d$ allows entropy to be directly calculated in the small-density limit (see extended fig.~\ref{efig:kinetics}), and this quantity bounds how much work the excited demon gas state could perform on an external system as it relaxes to an equilibrium state \cite{esposito_second_2011}.}

Kinetic theory further allows us to evaluate the rate of entropy production as \cite{Dorfman2021,Gaspard2022,Cercignani1994}
\begin{equation}
    \frac{dS}{dt} = \frac{k_B}{2}
    \int (g_+  - g_-)\ln{\frac{g_+}{g_-}}\,d\sigma \, d\bm{v}_*\,d\bm{v},
    \label{eq:schnakmain}
\end{equation}
from the Boltzmann equation, in which $g_{\pm}$ are the same as in eq.~\eqref{eq:col_op}.
From this equation, which is reminiscent of the Schnakenberg formula for systems described by a master equation~\cite{schnakenberg_network_1976},
it becomes apparent that the active compression of phase space ($\alpha$ in $g_+$) drives the system out of equilibrium.
The resulting entropy production rate for several diameter functions are shown in Fig.~\ref{fig:info_hydro}k as a function of mean excess free energy (i.e. excess entropy) $\langle\mathcal{F}-\mathcal{F}^\text{MB}\rangle$.
As shown in the figure, the entropy production rate is a monotonic function of excess entropy.

\noindent {\bf Controlling collective behavior through individual decisions.}
The velocity distribution for arbitrary diameter functions of velocity $D(\bm v)$ can be obtained by a semi-numerical method (see methods), as shown in fig.~\ref{fig:info_hydro}f-j.
The Maxwell-Boltzmann distribution (panel f) is deformed by non-constant diameter functions producing lower-entropy distribution functions with broken symmetries (g-j).
To obtain the nematic flocking state we have used a diameter function that breaks isotropy (Fig.~\ref{fig:info_hydro}j).
Breaking the four-fold rotational symmetry of the gas is sufficient to obtain a finite $\bm{Q}$ in the steady-state.
If we instead break parity symmetry in the velocity plane ($\bm v\to-\bm v$, Fig.~\ref{fig:info_hydro}i), a heat flux $\bm{q}\neq 0$ entering the equation of conservation of energy can arise in the steady state.
This can be predicted from the kinetic theory, which agrees well with numerical results obtained for small diameter changes (see Fig.~\ref{efig:kinetics}).
In the presence of a linear momentum source or sink (i.e.~another species of particle with different behavior) heat flux can be exploited to induce self-propulsion (see for example Fig.~\ref{efig:markov}).

\noindent \textbf{Hydrodynamic theory of informational flocking.} We now derive a hydrodynamic description of the flocking \dname\ gas (with diameter function described in Fig.~\ref{fig:nematic_intro}d-f) using a single relaxation time approximation~\cite{Bhatnagar1954} adapted to the demon gas.
In addition to the standard Navier-Stokes equations manifesting the conservation of mass, linear momentum, and energy (which are left unaffected), we find that the traceless part $Q_{i j} = P_{i j} - (P_{kk}/2)\delta_{ij}$ of the pressure tensor $P_{i j}$ evolves as
\begin{equation}
    \frac{D Q_{ij}}{D t} = \partial_k R_{ijk} + \frac{1}{\tau} \left[ 
      \frac{Q_{0}}{Q}  - 1
    \right]
    Q_{ij}
    \label{eom_Qtensor}
\end{equation}
in which $\tau$ is a relaxation time, $Q^2 \equiv \lVert \bm{Q} \rVert^2 \equiv Q_{\mu \nu} Q_{\nu \mu}/2 \geq 0$, $D/Dt$ is the material derivative, and
\begin{equation}
Q_0 = \frac{4 P_0}{\pi}\frac{\Delta D}{\Delta D + 2D_S}, \label{eq:dP_ssb}
\end{equation}
where $P_0 = \rho k_B T$
and $R_{ijk}$ is a higher-order current (see SI section \siNematicHydro\ for details).

In the homogeneous case, eq.~\eqref{eom_Qtensor} implies that the order parameter $Q$ quantifying the nematic flocking evolves as
\begin{equation}
    \partial_t Q  = \frac{1}{\tau}\left[ Q_0 - Q \right]
    \label{normQeom}
\end{equation}
while the direction of the principal axes of the symmetric tensor $\bm{Q}$ (corresponding to the nematic flocking direction) stays fixed~\footnote{In the homogeneous case, eq.~\eqref{eom_Qtensor} becomes $\partial_t Q_{ij} = [Q_0/Q - 1] Q_{ij}/\tau$. As $\bm{Q}$ is symmetric and traceless, it can be written as $\bm{Q} = x \bm{\sigma}_1 + y \bm{\sigma}_3$ in which $\bm{\sigma}_a$ are Pauli matrices.
Using polar coordinates $x + i y = Q e^{i \psi}$, we find $\partial_t Q = - V'(Q) = Q_0 - Q$ where $V(Q) = Q(Q - 2 Q_0)/2$ and $\partial_t \psi = 0$.}.
In the language of dynamical system theory, this describes a non-analytic version~\cite{Farutin2024} of a circle-pitchfork bifurcation~\cite{Kness1992} occurring at $Q_0 = 0$.
When $Q_0 > 0$, there is a circle of stable solutions $\bm{Q}$ with $Q=Q_0$ and an arbitrary orientation $\psi$~\footnote{When $Q_0>0$ there is also an unstable solution $\bm{Q} = 0$.
When $Q_0 \leq 0$, there is only a stable solution at $\bm{Q} = 0$, however $Q_0 < 0$ is unphysical in our case.}.

The steady-state solution of eq.~\eqref{normQeom} is shown in Fig.~\ref{fig:info_hydro}l and compared with numerical data.
The data collected for a gas with fixed anisotropy axis (explicitly broken symmetry, blue data) is nearly linear in $Q_0$, as predicted (note that $\langle |Q|\rangle$ does not approach $0$ as $Q_0\to 0$ due to thermal fluctuations).
The anisotropy of the gas is a weak function of temperature, remaining nearly constant as temperature is increased 50-fold (Fig.~\ref{fig:info_hydro}m).
Theory predicts that $Q/P_{kk}$ has no temperature dependence, thus observed trends in the simulations indicate a higher-order correction.
Simple arguments based on  coupling between nearby regions in the spatially extended gas, corresponding to adding diffusive terms in eq.~\eqref{eom_Qtensor} by linear response (e.g. $R_{\mu \nu \rho} \propto \partial_\rho Q_{\mu \nu})$, predict alignment of $\psi$ in the absence of fluctuations.

\changes{\noindent{\bf Measurement-induced flocking transition.} Going back to information thermodynamics, the steady-state magnitude of the order parameter $\bm{Q}$ (i.e., $Q_0$) is bounded by the quantity of information measured. 
Indeed, the work $W(Q_0)$ that the gas can exert on another system during relaxation to the Maxwell-Boltzmann distribution is bounded by $\Delta \mathcal{F}^*$ \cite{esposito_second_2011}, which in turn is bounded by $k_B T I$.
For the case of the flocking diameter function and to lowest order in activity ($\Delta D$), we find that $\Delta \mathcal{F}^* =\frac{1}{2} k_B T(\frac{\pi Q_0}{4 P_0})^2$. Hence, we get 





\begin{equation}
    \left(\frac{Q_0}{P_0}\right)
    ^2 \leq \frac{32}{
 \pi^2} \, I.
\label{bound}
\end{equation}

}


\changes{\noindent In other words, the magnitude $Q_0$ of the informational flocking order parameter is bounded by the information $I$ accumulated through continuous measurements.
This suggests viewing the transition to informational flocking as a classical measurement-induced phase transition \cite{Jin2022,Pizzi2022,Willsher2022,Pizzi2024,Gerbino2025}.}

\changes{
In quantum MIPTs \cite{Skinner2019,Choi2020,annurev}, local projective measurements may counteract the scrambling of information originating from the underlying unitary dynamics.
Here, the combination of local measurements and control actions may counteract the scrambling of information originating from the underlying chaotic dynamics (e.g. collisions in the gas).
Classically, the act of measurement itself does not affect the system\footnote{It can however affect the knowledge we have of a system \cite{Jin2022,Gerbino2025}.} so, as in Refs.~\cite{Pizzi2022,Willsher2022,Pizzi2024}, control actions are performed that modify the state of the system according to the outcome of the measurements.
In contrast with these works, here microscopic control actions are not necessarily designed to mimic quantum measurements, but include bio-inspired behavioral decisions.}

\noindent{\bf Exploiting non-thermal noise.} 
At scales beyond several $\mu$m, thermal noise will be too weak to agitate information-processing agents, while non-thermal noise sources (turbulence, vibration) are common.
Here, we use a pattern formation task to demonstrate that non-thermal noise can be used to produce low-entropy collective macrostates even in complex \changes{fluctuating} environments.
We consider two populations of particles (blue and red) that tend to move in opposite directions $\pm \bm{\xi}(\bm{x})$ set by a fixed field $\bm{\xi}(\bm{x})$.
The linear momentum conservation of our model does not preclude the formation of density patterns in single-species gases with diameter functions of position (see Fig.~\ref{efig:density_bump}), however momentum exchange between two species greatly enhances informational-self propulsion.

Using techniques from the field of policy optimization \cite{Schulman2017} we find that \dname\ gas propulsion is maximized for diameter functions of the form
\begin{equation}
    D(\bm{v},\bm{x}) = D_S + \Delta D \mathcal{H}\left(-\bm{v}\cdot\bm{\xi}(\bm{x})\right), \label{eq:step}
\end{equation}
\noindent where $\bm{\xi}$ is a vector field pointing along the direction of transport (see extended figure \ref{efig:learning} and SI section \siMD).
Regardless of noise ensemble (thermalized, mimicking a granular vibration table, or even a bath of self-propelled active particles) equation \ref{eq:step} is effective at producing robust flows.

\begin{figure*}[htbp]
\centering
\includegraphics[width=\textwidth,keepaspectratio]{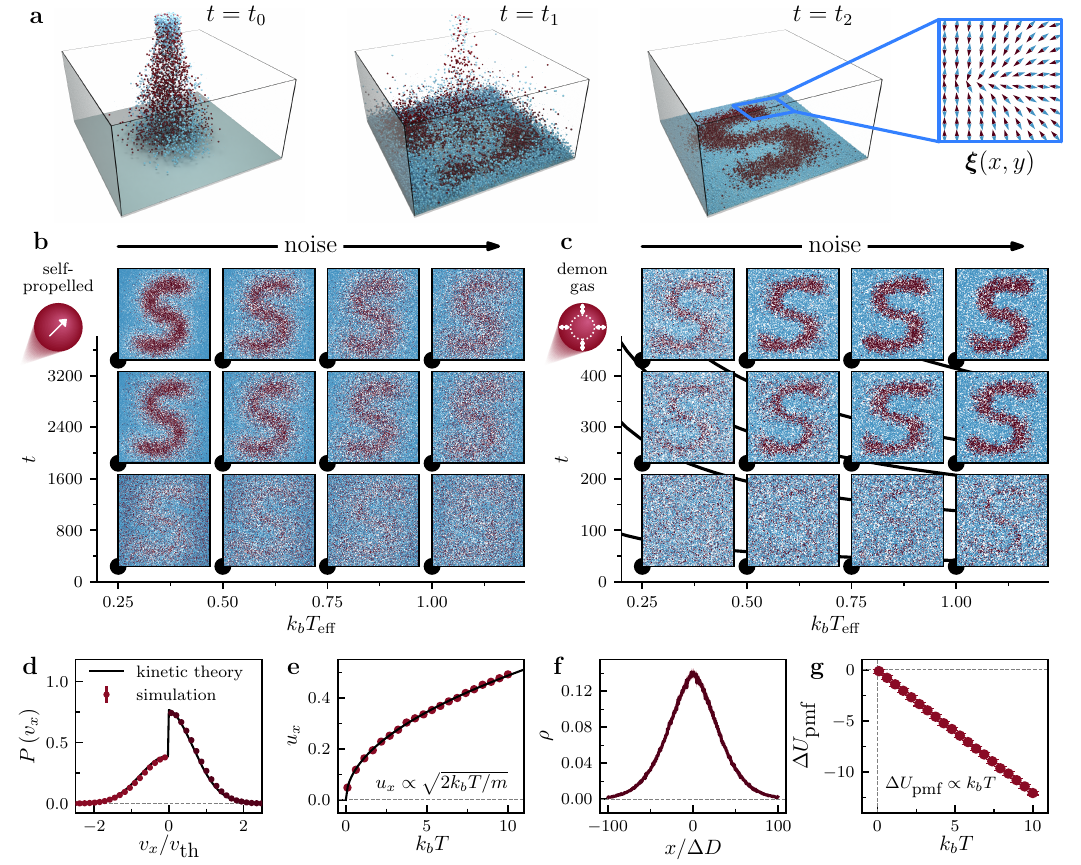}
\caption{
\textbf{Noise-driven active patterning.}
    \textbf{a}. A collection of variable-diameter particles are dropped into a hard-sided box.
    As particles undergo elastic collisions, their velocities are reduced by a linear drag ($\gamma=0.01$) term and they settle onto the bottom of the box.
    The two species (red and blue) have opposite $\bm{\xi}(\bm{x})$ fields (inset), driving separation into a designed pattern.
    \textbf{b}. Self-propelled particles that exert a constant magnitude force in the direction specified by $\bm{\xi}$ (i.e.~blue (red) particles push themselves away from (towards) the nearest segment of the pattern), while experiencing constant agitation (linear drag $\gamma=0.01$, random forces with mean magnitude $6k_bT_\textrm{eff}\gamma/\delta t$, see methods for details).
    As agitation is increased ($k_bT_{\text{eff}}$), pattern resolution degrades, since particles can only push with a fixed amount of force.
    Frames are instantaneous snapshots of simulations at the time and effective temperature indicated by the position of the inset lower left corner.
    \textbf{c}. Controller particles following the diameter rule of eq.~\ref{eq:step} and the same $\bm \xi$ field and non-thermal agitation as (b).
    The time required to obtain a given resolution decreases with increasing agitation.
    Black curves are $t = sL/\langle |\bm{v}| \rangle$ for $s=0.15\to 2$ where $L$ is the size of the simulation domain and $\langle | \bm{v}|\rangle = \sqrt{2k_bT_{\text{eff}}/m}$ is the mean speed of a particle in the gas.
    \textbf{d}. Velocity distribution function of \dname\ gas particles that adopt small (large) diameters when traveling right (left) immersed in a fixed-diameter isothermal gas. 
    Exchange of linear momentum with the passive gas allows the \dname\ gas to concentrate in the positive half of the velocity plane, as predicted by kinetic theory.
    \textbf{e}. Mean bulk velocity $u_x$ of the \dname\ gas in (d) as a function of temperature, with fixed $\Delta D=1$.
    Mean bulk flow due to collisional biasing is a fraction of the thermal speed.
    \textbf{f}. Density of \dname\ gas particles that adopt small (large) diameters when moving towards (away) from the origin.
    Selective collisions with each other and a surrounding passive isothermal gas concentrate them near the target location.
    \textbf{g}. Depth of an effective potential consistent with the increased density of \dname\ gas particles near the target in (f), as a function of temperature.
    }
\label{fig:active_patterning}
\end{figure*}

Figure \ref{fig:active_patterning}a and SI video V1 show the dynamics of two populations of informational particles dropped onto a surface.
Elastic collision of particles with the surface and each other in the presence of linear drag produces a non-thermal and non-stationary noise that the information-processing particles harness to create an S-shaped pattern.
Energy is dissipated via linear drag, eventually resulting in a stationary state.
This process can also be sustained for constantly agitated systems, see SI video V2.
Note that species separation can be observed in systems as small as $N=125$ particles (see extended Fig.~\ref{efig:small}).

In Fig.~\ref{fig:active_patterning}b-c, we compare patterns formed by continuously-agitated variable-diameter \dname\ particles and fixed-diameter particles with the ability to self-propel.
In both cases, particles are aware of $\bm{\xi}$, however in the case of propelled particles it is used to set the propulsion direction, equivalent to an externally applied potential $\nabla U(\bm x) = - f_w\bm \xi$.
The magnitude of their propulsive force, $f_w$, controls how far this system departs from equilibrium, whereas $\Delta D$ plays an analogous role for \dname\ particles.
Self-propelled particles demix (Fig.~\ref{fig:active_patterning}b) into the desired pattern.
However as non-thermal agitation is increased, the patterns become less and less visible because of the noise (expressible as an effective temperature $T_{\text{eff}}$).
In contrast, the resolution of patterns formed by \dname\ particles experiencing the same agitation (and following $\bm \xi$) appears to improve with additional noise (Fig.~\ref{fig:active_patterning}c).
In fact, the infinite-time resolution of \dname\ particles is entirely independent of the level of noise, while the time required to obtain a given resolution decreases with effective temperature, $\propto 1/\sqrt{T_{\text{eff}}}$.

The finite-time resolution is set by propulsion speed.
Returning to a thermal system in which a \dname\ gas is immersed in a fixed-diameter isothermal gas, kinetic theory (Fig.~\ref{fig:active_patterning}d-e) accurately predicts that the mean flow of the informational gas takes the form
\begin{equation}
    u_i=-\frac{m}{2 k_bT}\int v_i |\bm{v}|^2 f^d d\bm{v} = a(\rho,\Delta D) \sqrt{2k_bT/m}. \label{eq:drift_speed}
\end{equation}
\noindent The \dname\ gas effectively converts a temperature-independent fraction ($a(\rho,\Delta D)$) of the mean thermal kinetic energy into directed motion.

The noise-independence of patterns at long times arises due to the balancing of two effects which scale identically with noise.
Returning again to a thermal system of \dname\ gas particles, but now considering a diameter function of both velocity and location (Fig.~\ref{fig:active_patterning}f), we observe density gradients resulting from particle measurements.
An empirical effective potential, the so-called potential of mean force, $U_\text{pmf}(\bm{x})=-k_b T \ln \rho(\bm{x})/\rho_0$, can be computed to compare the system to one bound by an energetic potential \cite{Kirkwood1935}.
In Fig. \ref{fig:active_patterning}g the effective potential depth ($\Delta U_{\text{pmf}}$) is found to scale linearly with temperature, entirely compensating the effect of noise.

Similar effects occur in the non-thermal environment of Fig.~\ref{fig:active_patterning}c, and consequently increasing $k_bT_{\text{eff}}$ improves the short-time resolution of pattern-forming \dname\ gas particles, while having no effect on their eventual steady state.
Meanwhile, pattern formation under a temperature-independent driving force (e.g.~self-propulsion as in Fig.~\ref{fig:active_patterning}b) only degrades with additional noise.

\noindent{\bf Conclusion.}
Our results shed light on collective processes which exploit measurement and noise to produce non-equilibrium steady states in \changes{many-ratchet} systems.
Such processes are an important component of realistic systems where information processing and mechanical activity (e.g.~self propulsion) coexist.
\changes{Both types of activity are engines (i.e.~intermediaries between a non-equilibrated reservoir and a sink that generates work), however informational engines use their reservoir (i.e.~memory, prepared by dissipative erasure) to selectively couple to the momentarily `hot' flutuating environmental degrees of freedom.}
\changes{The conversion of noise into motive power requires control actions informed by measurements.
In practice, these feedback mechanisms will always convert energy stored in some fuel, thus blurring the theoretical distinction between informational and mechanical activity.
However at a fundamental level, measurements and the control actions they inform can be performed with arbitrarily small energy expenditure (zero, in the frictionless limit) as in the celebrated Feynman-Smoluchowski single-ratchet example \cite{Feynman2011}.
The only energy cost that is strictly required is the $k_{B} T \log{2}$ per bit associated with erasing the memory registry, as stipulated by Landauer's erasure principle \cite{Plenio2001}.}
 
\changes{Saturating Landauer's bound may be difficult in practice, however a key aspect of informational driving is that work output scales with the magnitude of environmental fluctuations, which we show can be exploited even for non-thermal reservoirs in which fluctuations can be very large.}
Hence, the \changes{practical} difference between these two flavors of non-equilibrium driving resembles the distinction between electrical rectifiers and batteries.
Informational driving, like a rectifier, produces bias from a fluctuating bias-free input, while a battery (like conventional self-propulsion) directly converts a chemical fuel into a bias.

The core concepts \changes{of informational activity investigated here by kinetic theory in classical many-ratchet }thermal systems and extended to non-thermal systems with reinforcement learning, may also apply to animal, bacterial, and micro-robotic swarms that \changes{collectively} seek to harness environmental noise to power goal-directed behaviors, \changes{and in which measurement-induced phase transitions similar to the measurement-induced flocking described in this work may also occur. This perspective may also be fruitful in the context of recently proposed active quantum systems \cite{rd46-hr3q,z3gm-32jn,Yamagishi2024,PhysRevResearch.4.013194,r6tm-nx19}.}


\noindent \textbf{Numerical model implementation}.
Molecular Dynamics (MD) simulations of the \dname\ gas were performed with the open-source software package \texttt{HOOMD-Blue} (v2.9.3) \cite{Anderson2020}.
All particles interact through a shifted Weeks-Chandler-Andersen potential (sWCA) \cite{WCA1971}, where the origin is shifted so that the radius of the particle is the potential's zero isoenergy surface.
This surface was modified by a custom updater in accordance with a diameter function.
Depending on the simulation, particles were simulated with velocity Verlet integration and a Nos\'{e}-Hoover thermostat or Langevin thermostat, as implemented in \texttt{HOOMD-blue}.
Nos\'e-Hoover simulations utilize a Verlet integration scheme to integrate dynamics governed by the extended Hamiltonian $\mathbb{H}=\sum_i \frac{\bf{p}_i^2}{2ms^2} + \frac{1}{2}\sum_{i,j,i\neq j}U(\bm{r}_i-\bm{r}_j) + \frac{p_s^2}{2\tau} + gk_bT\ln(s)$.
Here $U$ includes all (time-varying) interactions, $g$ is the number of momentum degrees of freedom, $(s,p_s)$ are extended coordinates associated with the thermostat, and $\tau$ is a coupling constant (in this study $\tau=0.1)$.
All data shown with the exception of Fig.~\ref{fig:active_patterning}a-c, \ref{efig:learning}d,f-h, and \ref{efig:small} (which employ Langevin dynamics) are integrated this way.
Langevin integration proceeds according to the equation of motion $m \dot{\bm{v}}=\bm{F}_\textrm{C} - \gamma\bm{v} + \bm{F}_\textrm{R}$, where $\bm{F}_\textrm{C}$ represents all external and (time-varying) interaction forces, $\bm{F}_\textrm{R}$ is a random force with average magnitude $2dk_bT\gamma/\delta t$ (for dimensions $d$ and timestep $\delta t$) and zero average.
To approximate the effect of vibration by a shaker table, in Fig.~\ref{fig:active_patterning}a-c, \ref{efig:learning}d, and \ref{efig:small} the vertical components of the random force $\bm{F}_\textrm{R}$ are drawn from a uniform distribution ten times wider than the horizontal components.
See SI section \siMD\ for details of each type of simulation employed.

\noindent \textbf{Free energy impulse fitting model}.
In Fig.~\ref{fig:info_hydro}b-e, the free energy change of a \dname\ gas after measurement is fit to a simple functional form to extract estimates of the amount of free energy injected per measurement event.
Here we outline the fitting procedure.
The free energy is computed from the (Shannon) entropy of a numerically sampled velocity distribution function (See SI section \siMD\ for numerical simulation details).
We subtract off the MB entropy as a baseline, and model the effect of measurement as a unidirectional coupling between the information-carrying degree of freedom (the diameters), and the velocity degrees of freedom.
We assume that the free energy of information carrying degree of freedom ($\mathcal{F}_\text{ic}$) receives impulses of strength $\mathcal{F}_m$ due to measurement which add to any elevated free energy remaining from the previous measurement ($\Delta\mathcal{F}^*$), and relaxes exponentially with a single timescale $\tau$,

\begin{align}
   \frac{d}{dt}\mathcal{F}_\text{ic} &= -\frac{1}{\tau}\mathcal{F}_\text{ic} \\
   \mathcal{F}_\text{ic} &= \left(\mathcal{F}_m + \Delta \mathcal{F}^*\right)e^{-\frac{t}{\tau}}
\end{align}

\noindent where we take $t=0$ as the time of measurement.
Relaxation of $\mathcal{F}_\text{ic}$ is assumed to occur by coupling to the velocity degrees of freedom ($\mathcal{F}$).
The velocity degrees of freedom are also assumed to exponentially relax to equilibrium (with the same timescale $\tau$),

\begin{align}
\frac{d}{dt}\mathcal{F} &= -\frac{d}{dt}\mathcal{F}_\text{ic} - \frac{1}{\tau} \mathcal{F} \\
\mathcal{F} &= \left(\Delta \mathcal{F}^* + \left(\mathcal{F}_m + \Delta \mathcal{F}^*\right)\frac{t}{\tau}\right)e^{-\frac{t}{\tau}} \label{eq:impulse_neqfe}
\end{align}

To fit these parameters, several simulated measurement intervals $t_m$ were observed and the resulting free energy curves are collapsed onto the range $[0,t_m]$.
The parameters $\mathcal{F}_m$, $\Delta \mathcal{F}^*$, and $\tau$ can be seen in extended Fig.~\ref{efig:parameter_fitting}a-b, and raw free energy curves along with their fit lines can be seen in panels d-l.
Bootstrap fitting error is shown in extended Fig.~\ref{efig:parameter_fitting}a-c.

\noindent \textbf{Informational propulsion}.
To compliment our analysis of the rate of information acquisition and departure from equilibrium in a homogeneous system of \dname\ particles (Fig.~\ref{fig:info_hydro}b-e), we also studied the simplified case of a single \dname\ particle immersed in a isothermal gas of fixed-diameter particles.
In this situation, scattering between the \dname\ particle and the passive gas allows for momentum transfer and spontaneous drift of the \dname\ particle (an effect not observable in homogeneous \dname\ systems with diameter functions of velocity only, due to linear momentum conservation).
The mean drift velocity achieved by the particle is a result of a balance between collisional drag against the passive thermal gas and biased scattering due to size changes.
In extended Fig.~\ref{efig:markov}a we construct a two-state Markov chain model from simple kinetic arguments, and in \ref{efig:markov}b we collapse all transitions to obtain the terms of a transition matrix describing the process.
In extended Fig.~\ref{efig:markov}c we compare the entropy rate of the process (calculated with eq.~\ref{eq:markov_entropy}) to the work required to drag the \dname\ particle through the isothermal gas at the drift velocity observed in simulation and find qualitative agreement, particularly with regards to the location of the maxima.
See SI section \siMarkov\ for additional details.

\noindent\textbf{Kinetic theory of variable diameter hard disks.}
Conservation of momentum and energy determine the velocities of a pair of hard disks after collision.
Assuming uniform particle masses,

\begin{align}
    \bm{v} + \bm{v}_* &= \bm{v}' + \bm{v}_*' \\
    |\bm{v}|^2 + |\bm{v}_*|^2 &= |\bm{v}'|^2 + |\bm{v}_*'|^2.
\end{align}

\noindent If $\hat{n}$ is a unit vector that points between disk centers at the point of contact (note that $\hat{n}$ therefore depends upon diameter), then the post-collision velocities can be written, 

\begin{align}
    \bm{v}' &= \bm{v} - ((\bm{v}-\bm{v}_*)\cdot\hat{n})\hat{n} \\
    \bm{v}_*' &= \bm{v}_* + ((\bm{v}-\bm{v}_*)\cdot\hat{n})\hat{n}.
\end{align}

\noindent For disks of fixed diameter, this scattering process is microreversible; $(\bm{v},\bm{v}_*)$ maps to $(\bm{v}',\bm{v}_*')$ and $(-\bm{v}',-\bm{v}_*')$ maps to $(-\bm{v},-\bm{v}_*)$, resulting in a Jacobian for the change of variables equal to $-1$ \cite{Cercignani1994}.
This one-to-one mapping allows all four velocity terms to be collected under a common unit disk integration in the typical quadratic Boltzmann collision operator (see SI section \siKinetic\ for additional details).
When particle diameters are instead functions of velocity, the microreversibility of the collision process is broken, but can be restored with a shift of relative positions.
In a coordinate frame centered on the point of contact for the forward collision, a shift in position of $\alpha = (D(\bm{v}')+D(\bm{v}_*'))/(D(\bm{v})+D(\bm{v}_*))$ along the reflection plane of the forward collision, applied to one particle during the reverse process restores the symmetry of velocity mapping (see extended Fig.~\ref{efig:microreversibility}).
Note that when \dname\ gas particles are immersed in a fixed-diameter gas, one set of diameters (starred or unstarred) is constant.

We now solve for the single particle probability distribution function ($f(\bm{v})$) that satisfies the steady state form of eq.~\ref{eq:boltzmann_equation} for the \dname\ gas collision operator by examining the collision invariants of the gas.
If we consider a moment of the time-varying distribution function $f(\bm{v},t)$ with a non-time-varying test function $\phi$:

\begin{equation}
    \langle \phi\rangle_f = \int f(\bm{v},t)\phi(\bm{v})d\bm{v}, \label{eq:test_func}
\end{equation}

\noindent the time evolution of this quantity is

\begin{equation}
    \frac{d}{dt} \langle \phi \rangle_f = \int \phi(\bm{v}) \mathcal{C}(f)(\bm{v},t) d\bm{v} \label{eq:time_evolution}.
\end{equation}

\noindent The function $\phi$ can be brought into the integrand of the collision operator since it is not a function of time.
As we are now considering an expression integrating over all $\bm{v}$ and $\bm{v}_*$ (and by extension $\bm{v}'$ and $\bm{v}_*'$), the choice of primed and starred variables is entirely arbitrary.
A typical procedure is to average over the exchange of velocity variables \cite{Cercignani1994}, producing the following for the case of $\alpha=1$,

\begin{equation}
    \frac{1}{8}\iiint (\phi+\phi_*-\phi'-\phi_*')\left( f'f_*' - ff_* \right) B(\sigma, \Delta \bm{v}) d\bm{v}_* d\bm{v} d\sigma,
\end{equation}

\noindent where $B$ is the collision kernel (see SI section \siKinetic) and $\sigma$ the impact parameter.
If $\phi$ is a quantity that is conserved through the collision, then it remains unchanged and $\frac{d}{dt}\langle \phi\rangle_f=0$.
This defines a summational collisional invariant for the system.
For the Boltzmann equation, summational invariants can only be linear combinations of the microscopically conserved quantities, $\mathcal{M} = 1,\, \bm{v},\, |\bm{v}|^2$, which are unchanged during collision by construction \cite{Cercignani1990}.
Therefore there exist constants $(a,\bm{b},c)$ that define any summational collision invariant:

\begin{equation}
    h(\bm{v}) = a + \bm{b}\cdot \bm{v} + c|\bm{v}|^2.
\end{equation} \label{eq:invariant_combo}

\noindent For the \dname\ gas with $\alpha\neq 1$, the same procedure of exchanging velocity variables allows a common term to be collected, $h=(D_\text{const}+D(\bm{v}))\phi$ for the mixed \dname-passive case, or $h=D(\bm{v})\phi$ more generally.
This term must satisfy the same condition $h+h_*-h'-h_*'=0$ for it to remain unchanged by collisions (i.e.~be an invariant).
Beginning with the MB velocity distribution function in $2$D,

\begin{equation}
    f^\textrm{MB}(\bm{v}) = \frac{m\rho}{2\pi kT}e^{-\frac{m|\bm{p}|^2}{2kT}},
\end{equation} \label{eq:max_boltz_dist}

\noindent where $\bm{p}=\bm{v}-\langle \bm{v} \rangle = \bm{v}-\bm{u}$ is the velocity of the gas in the flow frame, we express the \dname\ gas distribution function as $f^d = (1+\mathcal{D})f^\textrm{MB}$, where $\mathcal{D}$ is a function of velocity only.
We can treat the function $(1+\mathcal{D})$ as a test function operating on the MB distribution, with collisional invariant $h=D(1+\mathcal{D})$.
The four constants $a,b_x,b_y,c$ which define invariant $h$ (and therefore $\mathcal{D}$) can be found by requiring that the first four conserved moments ($\mathcal{M}$) of the \dname\ gas velocity distribution are equal to the first four moments of the equilibrium MB distribution,

\begin{align}
    \int \mathcal{M} f^\textrm{MB} d\bm{v} &= \int \mathcal{M} (1+\mathcal{D})f^\textrm{MB} d\bm{v} \nonumber \\
    0 &= \int \mathcal{M} \mathcal{D} f^\textrm{MB} d\bm{v},
\end{align}

\noindent i.e.~the \dname\ gas conserves number, momentum, and energy in all collisions.
Note that this restriction does not constrain the third-order moments which define the heat flux tensor, or the individual entries of the pressure tensor (derived from second-order moments).

From the expression for the \dname\ gas collision invariant one obtains $(1+\mathcal{D}) = h D^{-1}$.
Note that if $D = 0$ in some region around a point $\bm{v}_0$, then this region would become a sink for gas density and the entire gas would eventually come to reside there without further collisions.
Expressing the expectation value over $f^\textrm{MB}$ as $\langle \phi \rangle_m = \int \phi f^\textrm{MB} d\bm{v}$,

\begin{align}
    \langle \mathcal{M} \rangle_m & = \langle \mathcal{M}(1+\mathcal{D})\rangle_m \nonumber \\
    & = \langle \mathcal{M} h D^{-1}\rangle_m \nonumber \\
    & = a\langle \mathcal{M} D^{-1}\rangle_m + b_x \langle \mathcal{M} v_x D^{-1}\rangle_m + \nonumber \\
    &\;\;\;\;\;b_y \langle \mathcal{M} v_y D^{-1}\rangle_m +  c \langle \mathcal{M} |\bm{v}|^2 D^{-1}\rangle_m.
\end{align}

\noindent For each of the conserved moments ($\mathcal{M}=1,\, v_x,\, v_y,\, |\bm{v}|^2$) a new expression is generated, allowing for the definition of a full rank system of linear equations,

\begin{align}
\mathcal{A}\cdot G &= \mathcal{B} \\
\mathcal{A} &=\begin{bmatrix} \langle \frac{1}{D} \rangle_m && \langle \frac{v_x}{D} \rangle_m && \langle \frac{v_y}{D} \rangle_m && \langle \frac{|\bm{v}|^2}{D} \rangle_m \\
. && \langle \frac{v_x^2}{D}\rangle_m && \langle \frac{v_x v_y}{D}\rangle_m && \langle \frac{v_x |\bm{v}|^2}{D} \rangle_m \\
. && . && \langle \frac{v_y^2}{D} \rangle_m && \langle \frac{v_y|\bm{v}|^2}{D}\rangle_m \\
. && . && . && \langle \frac{|\bm{v}|^4}{D} \rangle_m
\end{bmatrix} \\
    G &= \begin{bmatrix} a && b_x && b_y && c \end{bmatrix}^\top \\
    \mathcal{B} &= \begin{bmatrix} \rho && \rho u_x && \rho u_y && \rho|\bm{u}|^2/2 + \rho kT
\end{bmatrix}^\top,
\end{align}

\noindent where the matrix $\mathcal{A}$ is symmetric and redundant entries are not shown.
Using the solution of this system of equations the \dname\ gas velocity distribution can be written as

\begin{equation}
    f^d = f^\textrm{MB}(1+\mathcal{D}) = f^\textrm{MB} D^{-1} h = f^\textrm{MB} D^{-1}\left(\bm{M} \cdot (\mathcal{A}^{-1}\mathcal{B}) \right),
\end{equation}

\noindent Where $\bm{M}=[1,v_x,v_y,|\bm{v}|^2]$.
For arbitrary diameter functions, the terms in the matrix $\mathcal{A}$ and its inverse are most easily found by numerical methods.
See the supplemental information for functional forms of $\mathcal{A}$ matrices for selected diameter functions.

Comparison between kinetic theory and simulation demonstrates strong agreement in the limit of small diameter changes and densities.
Extended Fig.~\ref{efig:kinetics}a-c compare a numerically simulated gas and properties derived from a semi-numerically obtained velocity distribution function.
While the departure of gas pressure from the MB distribution pressure is only quantitatively captured at small $\Delta D$, heat flux predictions remain accurate up to larger changes.
Extended Fig.~\ref{efig:kinetics}d compares a calculation of the mean drift speed (normalized by thermal speed $v_\text{th} = \sqrt{2k_bT/m}$) from kinetic theory (eq.~\ref{eq:drift_speed}) applied to a dilute mixed \dname-passive gas and numerical data, finding excellent agreement.
Finally, extended Fig.~\ref{efig:kinetics}ai-cv demonstrates how some collective properties of various \dname\ gases (computed from the kinetic theory) vary as $\Delta D$ is increased.

\noindent\textbf{Reinforcement learning for non-thermal environments}.
To explore the behavior of the \dname\ gas model subjected to non-thermal noise sources, we implemented an extension in which the diameter function was represented by a feed-forward neural network.
Using standard policy optimization algorithms \cite{Sutton2018, Schulman2017} (shown schematically in Fig.~\ref{efig:learning}a and detailed in SI section \siMD), the diameter function was optimized to bring particles to and keep them near the origin of mixed \dname-passive simulations.
Successful learning produced probabilistic diameter functions that drove \dname\ particles to accumulate at high density near their goal (Fig.~\ref{efig:learning}b).
Figure \ref{efig:learning}c-h shows the result of learning across a variety of simulation types (detailed in the SI).
In all cases, the dominant contribution to the diameter function took the form of eq.~\ref{eq:step}, with $\bm\xi(\bm x)$ pointing along the direction of least-time approach to the target location (white arrows).

\noindent\textbf{Spontaneous symmetry breaking.}
To obtain a closed equation for the growth of local pressure asymmetry, the linear, single relaxation time collision operator approximation is used,

\begin{equation}
\mathcal{C}_\textrm{lin} = \frac{1}{\tau}\left(f^d - f \right),
\end{equation}

\noindent where $f^d$ is the steady-state \dname\ gas velocity distribution and $\tau$ is the collision timescale.
The time evolution of the mean of any function of velocity can therefore be estimated as eq.~\ref{eq:test_func}, which produces equations of the form $d\langle \phi\rangle /dt = (\bar{\phi}-\langle \phi\rangle)/\tau$, where $\bar{\phi}$ is the mean value of the moment over the reference distribution $f_d$.
For the diameter function described in Fig.~\ref{fig:nematic_intro}d-f, $f_d$ can also be found analytically,

\begin{equation}
    f^d = \frac{a}{D(\theta)} f^\textrm{MB}
\end{equation}

\noindent where $f^\textrm{MB}$ is the Maxwellian distribution and $a=2D_S(D_S+\Delta D)/(2D_S+\Delta D)$.
See SI section \siNematicKinetic\ for details of this derivation, and SI section \siNematicHydro\ for derivation of the hydrodynamic relations for $\bm{Q}$.

\noindent\textbf{Data Availability}

\noindent The data underlying this article are available in a Zenodo repository, at https://dx.doi.org/[TBD]



\section*{Acknowledgments}
V.V.~acknowledges support from the Army Research Office under Grants No.~W911NF-22-2-0109 and No.~W911NF-23-1-0212, from the National Science Foundation under Grant No.~DMR-2118415 and from the Theory in Biology program of the Chan Zuckerberg Initiative. 
B.V.S.~acknowledges support from a MRSEC-funded (NSF DMR-2011864) Kadanoff-Rice fellowship.
M.F.~and V.V.~acknowledge support from the France Chicago center through a FACCTS grant.
This work was completed in part with resources provided by the University of Chicago Research Computing Center.
This research was partly supported by the National Science Foundation through the Physics Frontier Center for Living Systems (Grant No.~2317138) and the National Institute for the Theory and Mathematics in Biology (NITMB).
The authors would like to thank K.~Husain, D.~Martin and A.~Murugan for helpful conversations.

\section*{Author Contributions}
B.V.S., M.F., and V.V.~contributed to the design of the study and writing of the manuscript.
B.V.S.~ and M.F.~ performed analytical derivations.
B.V.S.~performed the numerical simulations.

\bibliography{main}

\setcounter{figure}{0}
\makeatletter 
\renewcommand{\thefigure}{E\@arabic\c@figure}
\makeatother

\begin{figure*}[htbp]
\centering
\includegraphics[width=\textwidth,keepaspectratio]{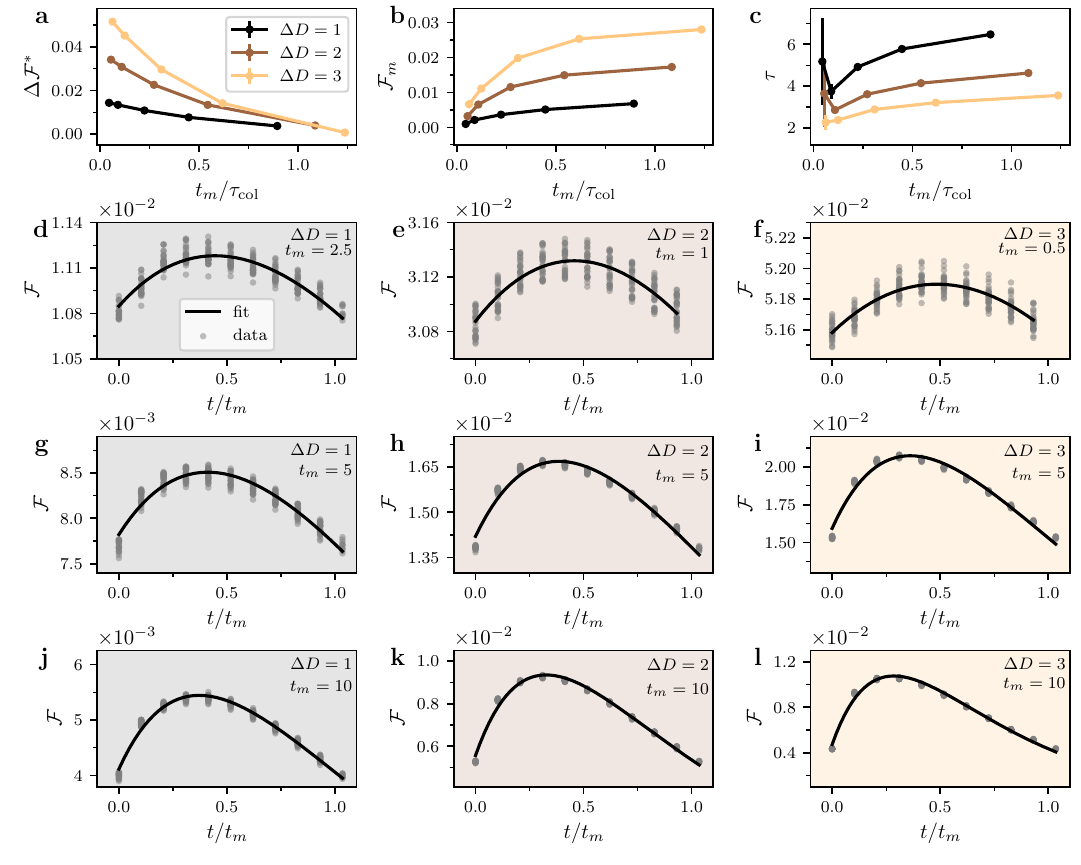}
\caption{
\textbf{Parameter fitting of the free energy of the \dname\ gas}.
\textbf{a}-\textbf{c} Estimated parameters of the free energy cycles of \dname\ gases with various measurement intervals ($t_m$) and diameter differences ($\Delta D$).
These parameters summarize the data presented in \textbf{d}-\textbf{l}.
\textbf{a}. Value of free energy at the start and end of a measurement cycle.
\textbf{b}. Free energy impulse delivered to the \dname\ gas from measurement and diameter change.
\textbf{c}. Relaxation parameter for free energy dynamics.
\textbf{d}-\textbf{l}. Free energy data, fit by the functional form presented in methods to extract the parameters in \textbf{a}-\textbf{c}.
Rows increase measurement time top to bottom, columns increase $\Delta D$ left to right.
Note that the smallest measurement time that can be reliably fit is a function of $\Delta D$.
}\label{efig:parameter_fitting}
\end{figure*}

\begin{figure*}[ht]
\centering
\includegraphics[width=\textwidth,keepaspectratio]{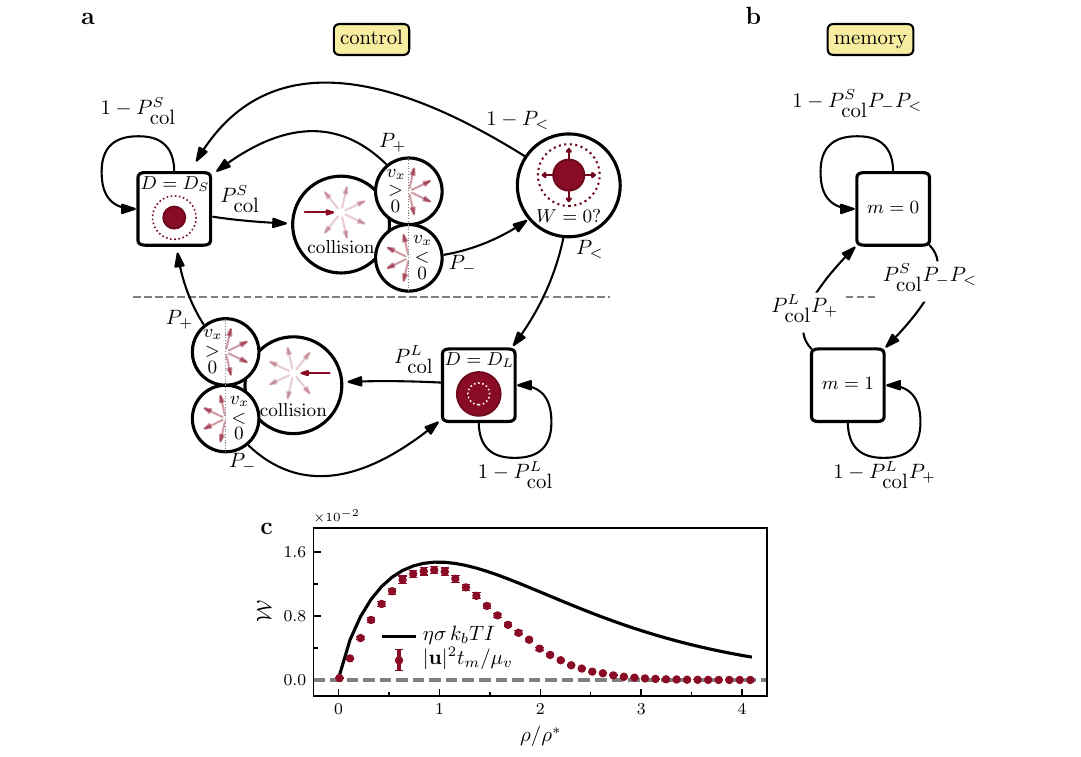}
\caption{
\textbf{Markov chain model and simulation of a \dname\ particle immersed in an isothermal gas}.
\textbf{a}. Transition diagram for the states of a \dname\ particle following a binary diameter rule: $D_L$ when $v_x<0$, $D_S$ when $v_x>0$.
Collisions occur in the time between measurements ($t_m$) with probability $P_\text{col}^L$ (for large particles or $P_\text{col}^S$ for small) and randomize velocities (scattering into negative or positive velocities with probabilities $P_-$ or $P_+$ respectively).
For hard disks, shrinking diameter never requires the \dname\ particle to exert work on the surrounding gas, but expansions are only possible if no other obstacles (i.e.~particles) are nearby (probability $P_<$).
\textbf{b}. Condensed two-state Markov chain model transition diagram for the one-bit measurement that \dname\ particles collect to determine which diameter state to adopt.
In the steady state, the measurement sequence $M=[m_t,m_{t+t_m},\dots]$ has a mean entropy per measurement, $H(M)$.
\textbf{c}. Work done to propel a single \dname\ particle through a passive gas per measurement cycle (duration $t_m$).
Power dissipated in \dname\ particle motion is estimated from simulation by finding the density-dependent drift velocity $\bm{u}$ and mobility coefficient $\mu_v$.
The black curve is the average free energy dissipated in deleting the \dname\ particle's memory of prior measurements, obtained from a Markov chain model of the measurement process (see SI section \siMarkov\ for additional details), where $\sigma$ is the intrinsic efficiency of rectification by collisional biasing and $\eta$ is a fitted constant of order one.
Numerical and theoretical estimations agree on the location of the work maxima at a density of $\rho=\rho^* = 1/A$, where $A$ is the excluded volume change as the \dname\ particle changes between its two diameter states.
}\label{efig:markov}
\end{figure*}

\begin{figure*}[ht]
\centering
\includegraphics[width=0.5\textwidth,keepaspectratio]{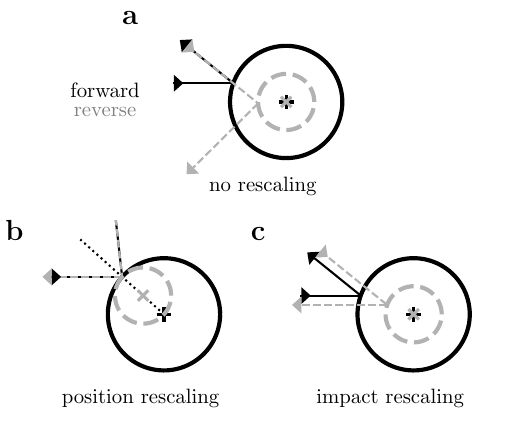}
\caption{
    \textbf{The microreversibility of hard disk collisions with velocity-dependent diameters.}
    Two particles of diameter $D(\bm{v})=D$ and $D(\bm{v}_*)=D_*$ with velocities $\bm{v}$ and $\bm{v}_*$ are shown as the collision of a point object (with velocity $\bm{v}_* - \bm{v}$) with a stationary disk of diameter $D+D_*$.
    \textbf{a}. In the coordinate frame centered on the stationary disk, the forward (black) and reverse (grey) collisions are not symmetric.
    \textbf{b}. In the coordinate frame centered on the point of contact forward and reverse collision processes are symmetric, provided that the stationary disk's position is rescaled by a factor $(D' + D_*')/(D + D_*)$.
    \textbf{c}. The symmetry of the collision can also be preserved in the frame centered on the stationary disk by rescaling the impact parameter of the incoming particle.
}\label{efig:microreversibility}
\end{figure*}

\begin{figure*}[ht]
\centering
\includegraphics[width=\textwidth,keepaspectratio]{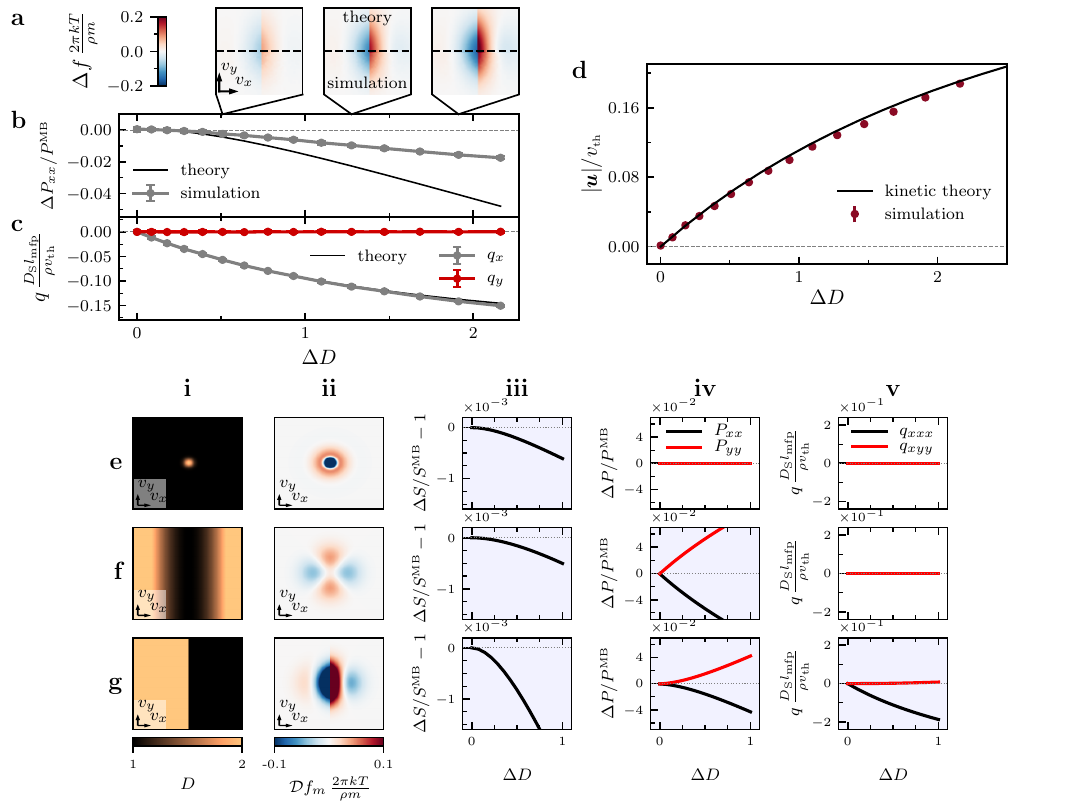}
\caption{
    \textbf{Observable consequences of \dname\ gas measurements.}
    {\bf a}. Comparison of \dname\ gas velocity distribution functions (relative to the MB distribution, $\Delta f = f^d - f^\textrm{MB}$) obtained by kinetic theory (top) and by MD simulation (bottom) at various diameter differences for a diameter step function.
    {\bf b}. Comparison of the pressure component (relative to MB, $\Delta P = P^d - P^\textrm{MB}$) aligned with the step function normal (denoted $x$) from kinetic theory and MD simulation.
    {\bf c}. Comparison of heat flux parallel and perpendicular to the step function normal.
    Heat flux is normalized by small diameter $D_\text{S}$, mean free path $l_\textrm{mfp}$, density $\rho$ and thermal speed $v_\textrm{th}$.
    {\bf d}. Comparison of the average speed of \dname\ particles immersed in an isothermal passive (fixed-diameter) gas, obtained by MD simulation and from kinetic theory.
    Heat flux generated by diameter change interacts with the passive species and results in net drift.
    \textbf{i}. Diameter in the velocity plane, \textbf{ii}. deviation from the MB distribution, \textbf{iii}. distribution entropy ($S=\int f^d\,\ln f^d d\bm{v}$) relative to the MB distribution, \textbf{iv}. pressure ($P_{ij}=\rho m\int v_iv_jf^d d\bm{v}$) relative to the MB distribution, and \textbf{v}. heat flux ($q_i = (\rho m/2)\int v_i |\bm{v}|^2 f^d d\bm{v}$) for \textbf{e}. a radially symmetric Gaussian diameter function, \textbf{f}. an anisotropic quadratic function, and \textbf{g}. a step function.
    }\label{efig:kinetics}
\end{figure*}

\begin{figure*}[htbp]
\centering
\includegraphics[width=0.5\textwidth,keepaspectratio]{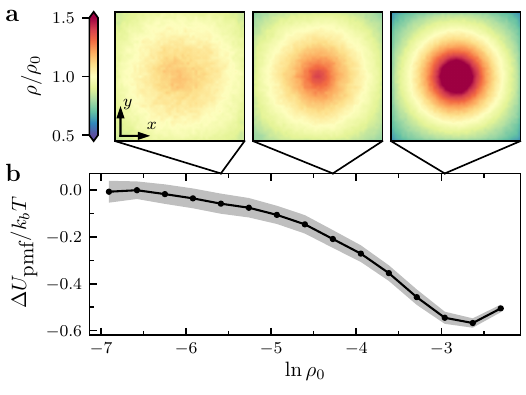}
\caption{
\textbf{Non-uniform density of a pure \dname\ gas.}
\textbf{a}. Snapshots of a demon gas with diameter function $D=D_S + \Delta D\mathcal{H}(-\bm{v}\cdot \bm{x})$.
Increasing initial density $\rho_0$ leads to a greater density enhancement near $\bm{x}=0$.
\textbf{b}. The maximum depth of the potential of mean force as initial density $\rho_0$ is increased.
Until $\ln\rho_0\approx -2.4$ increasing density enhances the depth of the potential well.
At greater densities crowding effects begin to inhibit \dname\ particle size changes.
}\label{efig:density_bump}
\end{figure*}

\begin{figure*}[htbp]
\centering
\includegraphics[width=\textwidth,keepaspectratio]{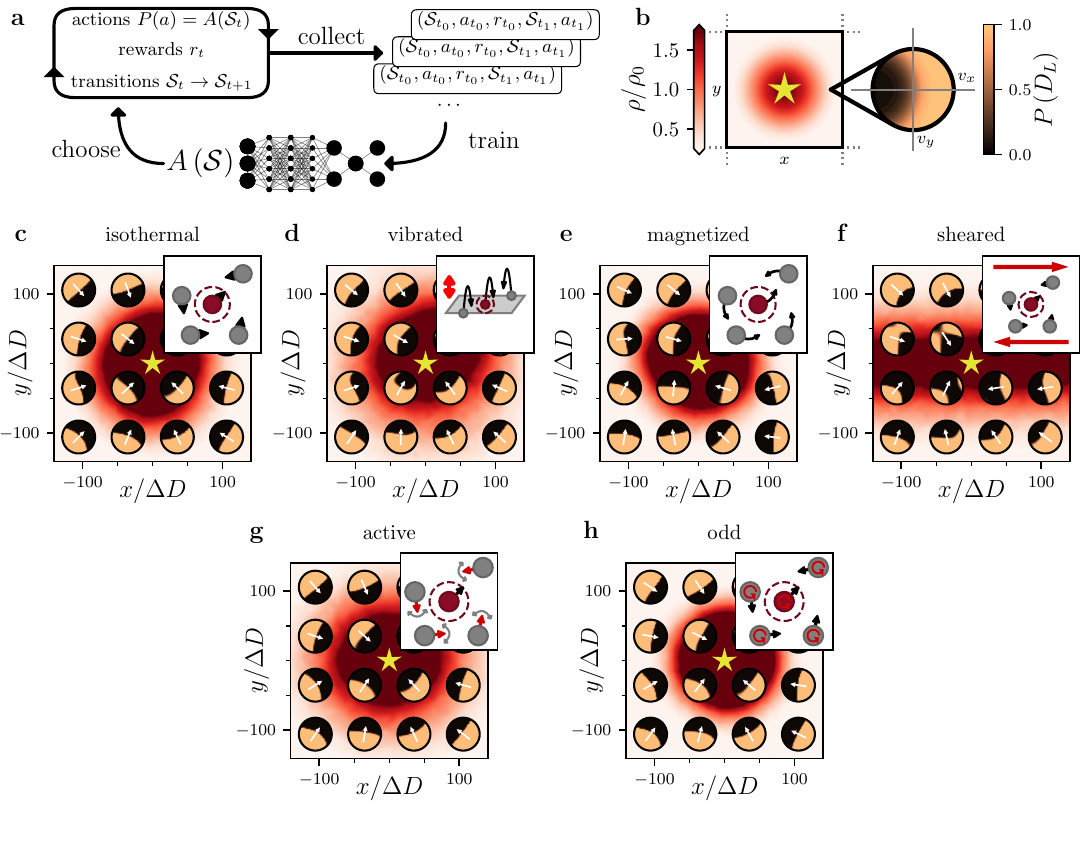}
\caption{
\textbf{Biasing strategies for \dname\ particles via reinforcement learning}.
\textbf{a}. The RL learning cycle.
Actions are chosen according to a policy function, and experiences of transitions are accumulated in a buffer.
This buffer is used to train a feed-forward artificial neural network that defines the policy function.
\textbf{b}. Locomotion task definition and legend.
Agents are embedded in a periodic environment with a singly-peaked (indicated by star) reward function.
Agent density after learning is shown in red.
The RL algorithm searches for a probabilistic discrete diameter function of $\bm{v},\bm{x}$ that most rapidly moves agents towards the maximum.
Plots in  \textbf{c}-\textbf{f} follow the same color scheme.
\textbf{c}-\textbf{h}. RL diameter functions under various transport physics.
Microscopic particle motions (black arrows) are schematically represented in insets.
In all cases, the vector field found by odd moments over the learned diameter distribution ($\bm{\xi}$) points along the shortest-time path towards the reward maximum (white arrows).
\textbf{c}. Isothermal environment.
\textbf{d}. Vibrated granular bed environment.
\textbf{e}. Isothermal dynamics with applied magnetic field.
\textbf{f}. Langevin dynamics with applied shear field.
\textbf{g}. Active gas bath.
Active particles have a fixed-magnitude force pointing along their rotationally-diffusing director.
\textbf{h}. Odd gas bath, in which all particles experience an equal and opposite transverse force during collisions, in addition to the usual hard core repulsion.
See SI section \siMD\ for various simulation details and parameters.}\label{efig:learning}
\end{figure*}

\begin{figure*}[htbp]
\centering
\includegraphics[width=\textwidth,keepaspectratio]{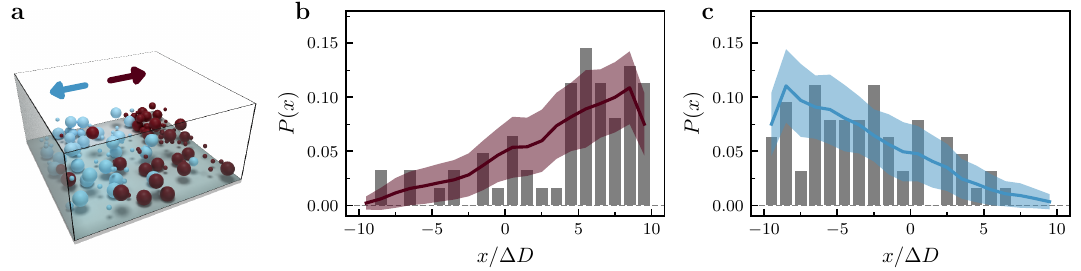}
\caption{
\textbf{Separation in a small vibrated \dname\ system}.
\textbf{a}. A small system of \dname\ particles ($N=125$) agitated by vertical forces and confined within a smooth-sided container.
Red (blue) particles attempt to move to positive (negative) $x$ positions.
\textbf{b}. Probability of finding red particles at a given $x$ location within the container.
Grey bars are data from one instant (as pictured in \textbf{a}), red data is the mean and one standard deviation calculated from 250 subsequent snapshots.
\textbf{c}. Data equivalent to \textbf{b} for blue particles.
Species separation is apparent even in a single snapshot.
See SI section \siMD\ for simulation details and parameters.}\label{efig:small}
\end{figure*}

\end{document}